\documentclass[9pt,article,twoside]{rmaa-rho}
\RMxAAtemplatetype{\RMxAA} % Select your article type
\vol{1}
\pages{1}
\thisyear{2026}
\doi{\href{https://www.astroscu.unam.mx/rmaa/RMxAA..XX-X}{https://www.astroscu.unam.mx/rmaa/RMxAA..XX-X}}

\title{Accretion in Binary Systems with Slow Stellar Winds}

\author[1]{Jesús A. Toalá\orcidlink{0000-0002-5406-0813}}
\author[2]{Emilio Tejeda\orcidlink{0000-0001-9936-6165}}
\author[1]{Raúl F. Maldonado\orcidlink{0000-0002-2236-7554}}
\author[1]{Diego A. Vasquez-Torres\orcidlink{0009-0008-2354-0049}}

\affil[1]{Instituto de Radioastronom\'{i}a y Astrof\'{i}sica, Universidad Nacional Autónoma de México, Morelia 58089, México}
\affil[2]{SECIHTI--Instituto de F\'{i}sica y Matem\'{a}ticas, Universidad Michoacana de San Nicol\'{a}s de Hidalgo, Ciudad Universitaria, 58040 Morelia, Mich., Mexico}
\leadauthor{J.~A.~Toalá et al.}
\smalltitle{Binary accretion in the Slow Wind Regime}

\corres{Jesús A. Toalá}
\email{j.toala@irya.unam.mx}

\received{\today}

\license{Texto de la licencia aquí}

\setbool{rho-abstract}{true} % Set false to hide the abstract
\setbool{rho-resumen}{true} % Set false to hide the abstract

\begin{abstract}
Wind accretion in binary systems is commonly described using the Bondi–Hoyle–Lyttleton (BHL) formalism. However, its standard implementation fails in the slow-wind regime, where the wind velocity of the donor star ($v_\mathrm{w}$) is comparable to or smaller than the orbital velocity of the accretor ($v_\mathrm{o}$). Tejeda \& Toalá recently proposed a geometrical correction to the BHL formalism that accounts for the wind aberration caused by the binary's orbital motion, which tilts the accretion cylinder and reduces its effective cross-section. Here we present a suite of smoothed particle hydrodynamic simulations performed with {\sc phantom} to test the geometric correction to the wind accretion in binary systems, with particular interest in systems operating in the slow-wind regime. We explore circular configurations and directly measure mass accretion efficiencies from the simulations. Our results confirm that the standard BHL prescription systematically overestimates accretion rates for $v_\mathrm{w}/v_\mathrm{o}<1$, while the geometrically corrected model agrees with the simulated efficiencies with remarkable accuracy. A key finding is that the velocity relevant for accretion estimates is not the value derived from the unperturbed stellar wind, but the local gas velocity measured upstream of the accretor. The gravitational potential of the accretor perturbs the flow, altering the effective relative velocity and modifying the accretion efficiency, particularly for compact orbits. These results provide strong numerical support for the geometrically corrected framework and establish a physically motivated basis for modeling wind-fed accretion in interacting binaries, including symbiotic systems.
\end{abstract}


\begin{resumen}
La acreción por viento en sistemas binarios se describe comúnmente mediante el formalismo de Bondi–Hoyle–Lyttleton (BHL). Sin embargo, su implementación estándar falla en el régimen de vientos lentos, donde la velocidad del viento de la estrella donadora ($v_\mathrm{w}$) es comparable o menor que la velocidad orbital del acretor ($v_\mathrm{o}$). Tejeda \& Toalá recientemente propusieron una corrección geométrica al formalismo BHL que representa la aberración del viento causada por el movimiento orbital de la binaria, la cual inclina el cilindro de acreción y reduce su sección eficaz efectiva. Aquí presentamos un conjunto de simulaciones hidrodinámicas de partículas suavizadas realizadas con {\sc phantom} para estudiar la corrección geométrica a la acreción por viento en sistemas binarios, con particular interes en sistemas en el régimen de viento lento. Exploramos configuraciones circulares y medimos directamente las eficiencias de acreción de masa a partir de las simulaciones. Nuestros resultados confirman que la prescripción estándar de BHL sobrestima sistemáticamente las tasas de acreción para $v_\mathrm{w}/v_\mathrm{o}<1$, mientras que el modelo geométricamente corregido reproduce las eficiencias obtenidas en las simulaciones con muy buena precisión. Un resultado clave es que la velocidad relevante para estimar la acreción no es la del viento estelar no perturbado, sino la velocidad local del gas medida corriente arriba en las inmediaciones del acretor. El potencial gravitacional del acretor perturba el flujo, alterando la velocidad relativa efectiva y modificando la eficiencia de acreción, particularmente en órbitas compactas. Estos resultados proporcionan un fuerte respaldo numérico al marco de acreción geométricamente corregido y establecen una base físicamente motivada para modelar la acreción alimentada por viento en binarias interactuantes, incluyendo sistemas simbióticos.
\end{resumen}

\begin{document}

\maketitle
\pagestyle{fancy}\thispagestyle{firststyle}

%----------------------------------------------------------

\section{Introduction}
\label{sec:intro}

Our understanding of wind accretion in binary systems has been built upon the classic Bondi-Hoyle-Lyttleton (BHL) scheme \citep[][]{HoyleLyttleton1939,Bondi1944}. Originally developed to describe a point mass accretor moving supersonically along a straight-line trajectory through a uniform medium, the BHL model found one of its earliest binary applications in the work of \citet{Davidson1973}, who used it to investigate the X-ray properties of high-mass X-ray binaries (HMXBs) such as Cen X-3 and Her X-1. Since those seminal studies, the BHL framework has become a fundamental tool for modeling wind-fed accretion in HMXBs \citep[e.g.,][and references therein]{Shakura2015,Weng2024}.

The BHL framework has also been widely applied to study symbiotic systems, where a white dwarf (WD) accretes material from the stellar wind of a late-type companion, typically a red giant or an asymptotic giant branch (AGB) star \citep{Mikolajewska2003,Merc2025}. However, this application has met with significantly less success \citep{Boffin2015}. Several studies have shown that for these configurations, the standard BHL prescription consistently and significantly overestimates the resulting mass accretion rate compared to numerical simulations \citep{Theuns1996,Nagae2004,Saladino2019}. In certain cases, the model even yields non-physical results where the predicted mass accretion efficiency exceeds unity.

The applicability of the BHL formalism in a binary context depends primarily on the relative velocity between the accretor and the donor's stellar wind. This relationship is characterized by the dimensionless velocity parameter 
\begin{equation}
 w = \frac{v_\mathrm{w}}{v_\mathrm{o}}   ,
 \label{eq:def_w}
\end{equation} 
where $v_\mathrm{w}$ is the wind velocity at the position of the accretor and $v_\mathrm{o}$ is the orbital velocity, with typical values of $v_\mathrm{o} \approx 5\text{--}50$ km s$^{-1}$. Fundamental to the BHL model is the ballistic approximation, which assumes the flow is sufficiently supersonic to neglect pressure gradients and treat fluid particles as being influenced solely by the accretor's gravitational attraction.

The direct implementation of the BHL model has proven highly successful for systems where $w > 1$, a condition commonly met in HMXBs where the winds of the donor stars are often at least one order of magnitude larger than the orbital velocities \citep[see, e.g.,][and references therein]{Hainich2020}.

In contrast, symbiotic systems typically operate in the slow-wind regime ($w \lesssim 1$), where the donor’s wind velocity is comparable to or smaller than the orbital velocity of the accretor. These systems feature low-velocity winds, with observed values in the range of $v_\mathrm{w} \approx 5\text{--}20$ km s$^{-1}$ \citep{Ramstedt2020,Wallstrom2025}. While the wind velocity in these cases may be comparable to or only slightly larger than the local speed of sound \citep[see, e.g., fig.~1 in][]{Maes2021}, the orbital velocity generally ensures that the net relative velocity remains supersonic, maintaining the formal validity of the ballistic approach. However, it is precisely in the $w \lesssim 1$ regime where the standard BHL prescription is known to systematically overestimate mass accretion rates.

To address the limitations of the standard BHL scheme in binary systems with low-velocity winds, an alternative mechanism known as Wind Roche Lobe Overflow (WRLO) was proposed in \citet{Mohamed2007} and \citet{Pod2007}. The WRLO scenario occurs when the wind injection radius ($R_1$)\footnote{In our numerical simulations, the wind injection radius corresponds exactly to the physical surface of the donor star where gas particles are introduced. In more complex physical models, this radius marks the onset of the dust condensation zone where radiation pressure begins to rapidly accelerate the wind.} is comparable to the donor's Roche lobe radius ($R_{\mathrm{RL}}$). In this regime, the slow-moving wind is effectively channeled toward the companion through the inner Lagrangian ($L_1$) point, leading to significantly higher accretion efficiencies than those predicted by standard isotropic wind models. However, \citet{Mohamed2012} later showed that if the wind injection radius remains smaller than $R_\mathrm{RL}$, the wind accretion regime is recovered regardless of whether the $w < 1$ condition is met.

Other attempts to mitigate the discrepancies of the BHL model in the slow-wind regime have often introduced an ad hoc efficiency parameter, $\alpha_{\mathrm{BHL}}$, typically ranging from $0.5$ to $0.8$ \citep[e.g.,][]{SaladinoPols2019,Lee2022,Malfait2024}. Alternatively, some studies circumvent these limitations by restricting their analysis to symbiotic systems where $w > 1$ \citep[e.g.,][]{Vathachira2026}.

A recent re-evaluation of the standard wind accretion model applied to binaries was presented by \cite{TejedaToala2025}.  In that work, our group demonstrated that introducing a simple geometrical correction to the standard BHL formulation is sufficient to alleviate the long-standing discrepancies found in systems operating in the slow-wind regime. This approach accounts for the fact that the stellar wind meets the accretor at a specific angle relative to its orbital motion, modifying the effective cross-section of the accretion cylinder.  This new prescription predicts reduced accretion efficiencies for $w < 1$, which are better aligned with numerical results, while recovering the standard BHL result for $w > 1$.

The geometrically corrected framework of \cite{TejedaToala2025} has already been applied to investigate specific objects, such as the symbiotic system R Aquarii \citep{VasquezTorresRAquarii}, and to model the long-term evolution of both interacting binaries \citep{Maldonado2025,Maldonado2025_WRLO} and planetary systems around evolved stars \citep{PadillaLopez2026}. Despite this progress, direct comparisons between this analytical prescription and hydrodynamical simulations available in the literature remain challenging. A primary difficulty lies in the inconsistent wind velocity profiles adopted across different studies, which complicates the benchmarking of accretion models. Moreover, the gravitational potential of the binary interaction substantially modifies the wind velocity experienced by the accretor relative to the unperturbed stellar wind. This effect is further complicated if the accretor is located within the wind acceleration zone of the primary star.

In this paper, we address these challenges by presenting a suite of hydrodynamical simulations to test wind accretion, with particular interest in the slow-wind regime ($w < 1$). However, we note that simulations in the $w > 1$ regime are also presented for comparison and discussion. We explore various circular configurations to directly measure local gas velocities and mass accretion efficiencies. These results are then compared with the predictions of both the standard BHL model and the geometrically corrected framework of \cite{TejedaToala2025}.

The remainder of this article is organized as follows: Section~\ref{sec:model} introduces the main model parameters. Section~\ref{sec:sim} details the numerical setup of our hydrodynamical simulations. Section~\ref{sec:results} presents the simulation results, which are subsequently discussed in Section~\ref{sec:discussion}. Finally, we summarize our primary findings in Section~\ref{sec:summary}.

\section{Model description}
\label{sec:model}

We consider a binary system in circular orbit consisting of a donor star with mass $m_{1}$ and an accreting companion with mass $m_{2}$. The system is characterized by its orbital separation $a$ with an orbital velocity $v_\mathrm{o}$ given by
\begin{equation}
    v_\mathrm{o} = \sqrt{\frac{G (m_1+m_2)}{a}}.
\end{equation}

The mass transfer process is governed by the interaction between the donor's stellar wind, moving at velocity $v_\mathrm{w}$ at the accretor's position, and the orbital motion of the secondary. This relationship is quantified by the dimensionless velocity parameter $w$ introduced in Equation~(\ref{eq:def_w}) and the reduced mass ratio $q$,
\begin{equation}
    q = \frac{m_2}{m_1 + m_2}.
    \label{eq:def_q} 
\end{equation}

The primary metric of our study is the mass accretion efficiency $\eta$, defined as
\begin{equation}
    \eta = \frac{\dot{M}_\mathrm{acc}}{\dot{M}_\mathrm{w}},
\end{equation}
where $\dot{M}_\mathrm{acc}$ is the mass accretion rate onto the secondary and $\dot{M}_\mathrm{w}$ the total mass-loss rate of the donor; both taken as positive quantities. Under this definition, the mass accretion efficiency cannot exceed a value of 1.

We compare the mass accretion efficiency obtained from the simulations ($\eta_\mathrm{sim}$) against two analytical models:
\begin{enumerate}
    \item \textbf{Standard BHL} ($\eta_\mathrm{BHL}$). The direct implementation of the BHL model to binary systems leads to
\begin{equation}
    \eta_\mathrm{BHL} = \frac{q^2}{w (1 + w^2)^{3/2}},
    \label{eq:eta_BHL}
\end{equation}
    \item \textbf{Geometrically corrected model} ($\eta_\mathrm{TT}$). The formulation by \cite{TejedaToala2025}, which accounts for the effective cross-section of the accretion cylinder by considering the non-zero angle between the wind and the orbital motion
\begin{equation}
\eta_\mathrm{TT} = \left(\frac{q}{1+w^2}\right)^2.
\label{eq:eta_TT}
\end{equation}    
\end{enumerate}

The complex hydrodynamical structures arising from these interactions have been extensively documented for decades in numerical studies \citep[see, e.g.,][]{Theuns1996,Nagae2004,deValBorro2009,HuarteEspinosa2013,Liu2017,Saladino2019,SaladinoPols2019,Lee2022,Malfait2024}. They clearly show the formation of an accretion disc around the secondary and a spiral-shaped accretion wake that characterizes wind-fed accretion in the slow-wind regime.

Such works highlight the need for a robust numerical approach to accurately capture the flow dynamics near the accretor. To achieve this, we employ three-dimensional hydrodynamical simulations to bridge the gap between the simplified analytical frameworks and the distorted velocity fields observed in our modeled systems. 

We remark here that our  hydrodynamical simulations are not meant to expand further our knowledge of the complex structures formed around the accreting binary systems. The simulations will be used to measure the mass accretion rate on the accretor, which will be ultimately compared with analytical predictions.

\section{Simulations}
\label{sec:sim}

We performed three-dimensional smoothed particle hydrodynamic (SPH) simulations using the {\sc phantom} code \citep{Price2018}.\footnote{\url{https://phantomsph.github.io/}} The simulations adopt an adiabatic equation of state of the form
\begin{equation}
    P = (\gamma - 1) \rho u,
\end{equation}
\noindent where $P$ is the gas pressure, $\rho$ is the volumetric density, $u$ is the internal energy, and $\gamma$ is the polytropic index. The temperature of the gas is calculated by adopting the ideal gas equation
\begin{equation}
    T = \frac{\mu m_\mathrm{H} }{k_\mathrm{B}}\frac{P}{\rho},
\end{equation}
\noindent where $\mu$ is the mean molecular weight, $m_\mathrm{H}$ is the hydrogen mass, and $k_\mathrm{B}$ is the Boltzmann's constant. Constant values for the polytropic index and the mean molecular weight are adopted as $\gamma=1.2$ and $\mu=1.26$; the choice of these values  are consistent for atomic material with typical temperatures of AGB  \citep{Millar2004,Maes2021}. Our simulations include H~I cooling following the methodology presented by \citet{Malfait2024}, where the cooling is calculated using the equation \citep[see][]{Spizer1978}
\begin{equation}
    \Lambda_\mathrm{HI} = 7.3\times10^{-19} \frac{n_\mathrm{e} n_\mathrm{H} e^{-118400/T}}{\rho}~\mathrm{erg}~\mathrm{cm}^3~\mathrm{s}^{-1},
\end{equation}
\noindent with $n_\mathrm{e}$ and $n_\mathrm{H}$ representing the number density of electrons and hydrogen, respectively. The binary simulations were performed adopting the wind routines predefined in {\sc phantom} and described in \citet{Siess2022}. 

{\sc phantom} does not include the complex mechanism of dust-driven winds. Instead, it adopts the free-wind approximation to artificially balance the gravity of the AGB star. An approximation that has proven to be reasonable for the binary configurations similar to those studied here \citep[e.g.,][]{Maes2021,Malfait2024,Esseldeurs2023}.

\begin{figure}
\begin{center}
\includegraphics[width=\linewidth]{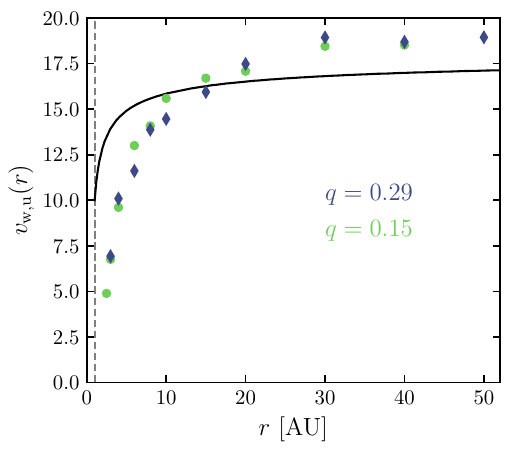}
\end{center}
\captionsetup{justification=justified}
\caption{Unperturbed wind velocity profile ($v_\mathrm{w,u}$) for the mass-losing donor star ($m_1$) computed by {\sc phantom} (solid line). This profile remains identical across all simulations in our suite. The wind is launched from the donor surface with an initial velocity $v_\mathrm{ini}=10$ km~s$^{-1}$, reaching a terminal speed of $v_{\infty}\lesssim 17.8$ km~s$^{-1}$. The dashed vertical line at $r=1$ AU indicates the wind injection radius ($R_1$). Symbols represent the local wind velocity ($v_\mathrm{w}$) measured upstream near the accretor for each individual simulation.}
\label{fig:wind1D}
\end{figure}

Our modeled systems are designed to reflect the general observable properties of known symbiotic systems.\footnote{The characteristics of a sample of symbiotic systems are listed in Appendix~\ref{sec:app}. We note that although some of those systems are eccentric, the simulations presented here are meant to describe their average mass accretion efficiencies as discussed in Section~\ref{sec:SySts}.} To explore the parameter space, we performed twenty SPH simulations divided into two distinct families, labeled A and B, based on the reduced mass ratio $q$ (see further details in Table\ref{tab:set1}):

\begin{itemize}
    \item \textbf{Family A} (Simulations (A3-A50): This set adopts a donor mass of $m_1 = 1.5$ M$_\odot$ and an accreting companion of $m_2 = 0.6$ M$_\odot$. These masses yield a fixed reduced mass ratio of $q \simeq 0.29$. We explored ten different orbital separations ranging from 3 to 50 AU. The labels refer to the adopted orbital separation, so that, for example, simulation A6 is from family A with an orbital separation of 6~AU.   
    
    \item \textbf{Family B} (Simulations B2.5-B40): This set adopts a donor mass of $m_1 = 1.7$ M$_\odot$ and an accreting companion of $m_2 = 0.3$ M$_\odot$. These masses yield a fixed reduced mass ratio of $q = 0.15$. We performed ten simulations with orbital separations ranging from 2.5 to 40 AU. The labeling scheme is the same as that described for simulations in family A.
\end{itemize}

In both families, the donor star properties are approximated to standard values for an AGB star, with a luminosity of $L_1 = 5315$ L$_\odot$ and a radius of $R_1 = 214$ R$_\odot$ ($\approx 1$ AU), the later consistent with the parameters of an AGB star \citep{Vassiliadis1993,McDonald2018}. The stellar wind is modeled by injecting SPH particles at a radial distance $r = R_1$ from the donor's center, using a constant initial launching velocity of $v_{\text{ini}} = 10$ km s$^{-1}$. The resolution parameter (\texttt{iwind\_res}) is  set to 8, which results in an average of $5\times10^{5}$ particles in the wind. The unperturbed wind velocity profile of the mass donor star computed by {\sc phantom} is presented as a solid line in Fig.~\ref{fig:wind1D}, which has a terminal wind velocity of $v_{\infty}\sim 17.8$ km~s$^{-1}$. The secondary (the accretor) is modeled as a sink particle with an accretion radius of 0.01 AU.

\begin{table*}
\begin{center}
\caption{Details of circular simulations analyzed in this paper. Note that $v_\mathrm{w}$ is the measured relative velocity of the gas with respect to the accretor, which is used to estimate $w$, $\eta_\mathrm{TT}$, and $\eta_\mathrm{BHL}$.}
%\footnotesize
\setlength{\tabcolsep}{\tabcolsep}  
%\scriptsize
\begin{tabular}{cccccccccccccc} 
\hline
$m_1$ & $m_2$ & $q$ & Label & $a$  &  $R_\mathrm{RL}$ & $v_\mathrm{o}$ & $v_\mathrm{w}$ & $w$ & $\eta_\mathrm{sim}$ & $\eta_\mathrm{TT}$ & $\eta_\mathrm{BHL}$ \\
  & & & & (AU)  & (AU)  & (km~s$^{-1}$) & (km~s$^{-1}$) & & & \\  
\hline
1.5 & 0.6 & 0.29 & A3  & 3.0  & 1.38  & 24.92 & 6.93 & 0.278 & 6.22$\times10^{-2}$ & 7.03$\times10^{-2}$ & 2.63$\times10^{-1}$ \\
& & & A4  & 4.0  & 1.84  & 21.58 & 10.09 & 0.468 & 5.14$\times10^{-2}$ & 5.49$\times10^{-2}$ & 1.29$\times10^{-1}$ \\
& & & A6  & 6.0  & 2.76  & 17.63 & 11.61 & 0.659 & 3.47$\times10^{-2}$ & 3.97$\times10^{-2}$ & 7.22$\times10^{-2}$ \\
& & & A8  & 8.0  & 3.68  & 15.27 & 13.87 & 0.908 & 2.39$\times10^{-2}$ & 2.45$\times10^{-2}$ & 3.65$\times10^{-2}$ \\
& & & A10  & 10.0 & 4.60  & 13.66 & 14.46 & 1.058 & 1.79$\times10^{-2}$ & 1.81$\times10^{-2}$ & 2.49$\times10^{-2}$ \\
& & & A15  & 15.0 & 6.90  & 11.15 & 15.94 & 1.429 & 0.86$\times10^{-2}$ & 0.88$\times10^{-2}$ & 1.07$\times10^{-2}$ \\
& & & A20  & 20.0 & 9.20  & 9.66  & 17.48 & 1.809 & 0.50$\times10^{-2}$ & 0.45$\times10^{-2}$ & 0.51$\times10^{-2}$ \\
& & & A30  & 30.0 & 13.79 & 7.89  & 18.93 & 2.399 & 2.25$\times10^{-3}$ & 1.78$\times10^{-3}$ & 1.93$\times10^{-3}$ \\
& & & A40  & 40.0 & 18.39 & 6.83  & 18.69 & 2.736 & 1.24$\times10^{-3}$ & 1.13$\times10^{-3}$ & 1.20$\times10^{-3}$ \\
& & & A50 & 50.0 & 22.99 & 6.11  & 18.94 & 3.099 & 7.72$\times10^{-4}$ & 7.26$\times10^{-4}$ & 7.63$\times10^{-4}$ \\
\hline
1.7 & 0.3 &0.15 & B2.5 & 2.5  & 1.33 & 26.66 & 4.89 & 0.183 & 2.51$\times10^{-2}$ & 2.11$\times10^{-2}$ & 1.17$\times10^{-1}$ \\
    &     &     & B3 & 3.0  & 1.59 & 24.34 & 6.76 & 0.277 & 2.10$\times10^{-2}$ & 1.94$\times10^{-2}$ & 7.24$\times10^{-2}$ \\
    &     &     & B4 & 4.0  & 2.13 & 21.08 & 9.61 & 0.455 & 1.53$\times10^{-2}$ & 1.54$\times10^{-2}$ & 3.72$\times10^{-2}$ \\
    &     &     & B6 & 6.0  & 3.19 & 17.21 & 13.00 & 0.755 & 9.20$\times10^{-3}$ & 9.12$\times10^{-3}$ & 1.51$\times10^{-2}$ \\
    &     &     & B8 & 8.0  & 4.25 & 13.79 & 14.08 & 1.021 & 5.85$\times10^{-3}$ & 5.39$\times10^{-3}$ & 7.55$\times10^{-3}$ \\
    &     &     & B10 & 10.0 & 5.31 & 13.33 & 15.59 & 1.169 & 3.99$\times10^{-3}$ & 4.01$\times10^{-3}$ & 5.28$\times10^{-3}$ \\
    &     &     & B15 & 15.0 & 7.97 & 10.88 & 16.70 & 1.535 & 1.90$\times10^{-3}$ & 2.00$\times10^{-3}$ & 2.38$\times10^{-3}$ \\
    &     &     & B20 & 20.0 & 10.63 & 9.43 & 17.08 & 1.811 & 1.07$\times10^{-3}$ & 1.22$\times10^{-3}$ & 1.40$\times10^{-3}$ \\
    &     &     & B30 & 30.0 & 15.94 & 7.70 & 18.45 & 2.396 & 4.72$\times10^{-4}$ & 4.95$\times10^{-4}$ & 5.37$\times10^{-4}$ \\
    &     &     & B40 & 40.0 & 21.26 & 6.67 & 18.52 & 2.777 & 2.22$\times10^{-4}$ & 2.96$\times10^{-4}$ & 3.15$\times10^{-4}$ \\
\hline
\end{tabular}
\label{tab:set1}
\end{center}
\end{table*}

\begin{figure}[t]
\begin{center}
\includegraphics[width=\linewidth]{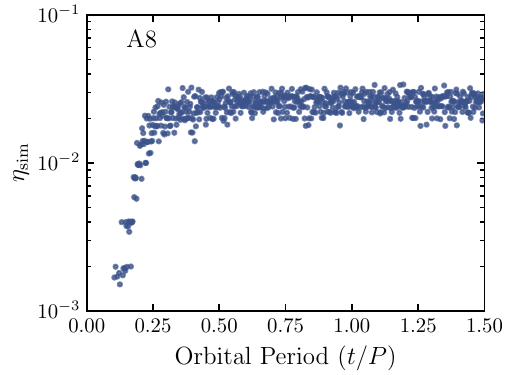}
\end{center}
\captionsetup{justification=justified}
\caption{Evolution of the mass accretion efficiency during the first one and a half orbital period resulted from simulation A8.}
\label{fig:eta_sim8}
\end{figure}  

\begin{figure*}
\begin{center}
\includegraphics[width=\linewidth]{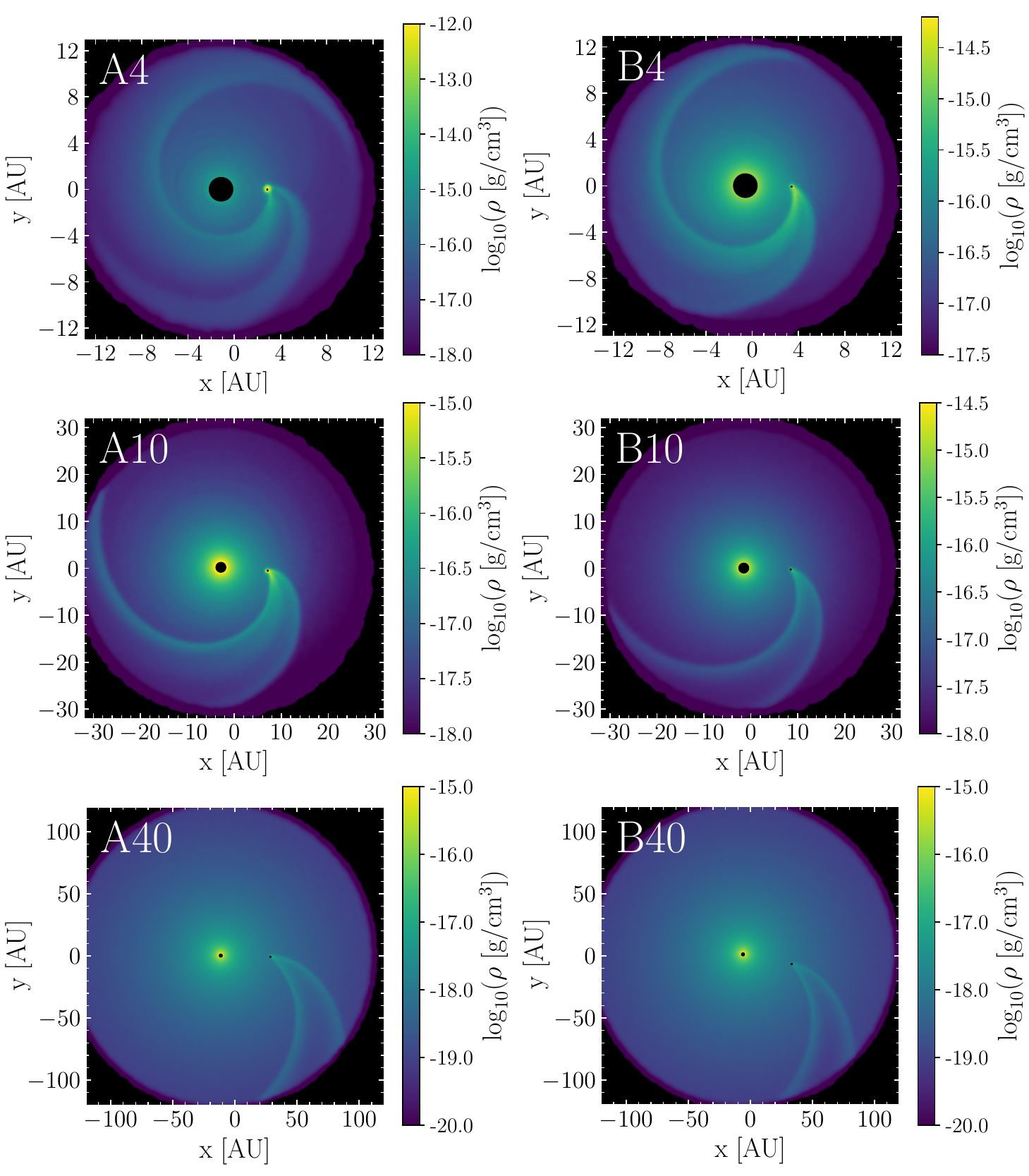}
\end{center}
\captionsetup{justification=justified}
\caption{Mid-plane volume density ($\rho$) maps for six representative three-dimensional simulations. The spiral-shaped wakes formed as a result of the binary interaction are clearly visible.}
\label{fig:all_sim}
\end{figure*}

Because the modeled systems span a range of binary separations, the physical dimensions of the numerical domains vary accordingly. To standardize the setup, the total computational box size for each simulation was scaled to three times the orbital separation. For example, a system with an orbital separation of $a = 4$~AU was modeled using a domain radius of 12~AU.

All simulations are performed in the inertial frame of the center of mass of the binary system and are allowed to evolve for at least 5 orbital periods, which is sufficient time for the accretor to reach a stable mass accretion rate. In the most extended cases, this steady state is achieved before completing even a single orbital period. {\sc phantom} reports the mass accretion efficiency directly from the simulation at each time step. For example, Fig.~\ref{fig:eta_sim8} shows the evolution of the mass accretion efficiency ($\eta_\mathrm{sim}$) from simulation A8. Table~\ref{tab:set1} reports the median value of $\eta_\mathrm{sim}$  obtained for all simulations.

We then extract the local wind velocity ($v_\text{w}$) by taking a time average over one full binary orbit to minimize numerical noise. The specific physical parameters and the resulting values for each individual run are summarized in Table~\ref{tab:set1}.

Finally, it is important to highlight here that our simulations are meant to explore the slow wind regime without entering the WRLO accretion regime. To confirm this, we calculate the Roche lobe radius ($r_\mathrm{RL}$) around the donor star, as listed in the 6th column of Table~\ref{tab:set1}. The table shows that in all cases, $r_\mathrm{RL} > R_1$, ensuring that we are modeling strictly wind accretion cases.

\section{Results}
\label{sec:results}

In agreement with previous numerical studies of accreting binaries, our simulations result in the formation of large-scale spiral-shaped wakes in addition to the accretion discs \citep[see, for example,][]{deValBorro2009,HuarteEspinosa2013,Lee2022,Malfait2024,Nagae2004,Saladino2019}. Fig.~\ref{fig:all_sim} presents the large-scale gas density distribution for a representative subset of six simulations.

\begin{figure*}[t]
\begin{center}
\includegraphics[width=\linewidth]{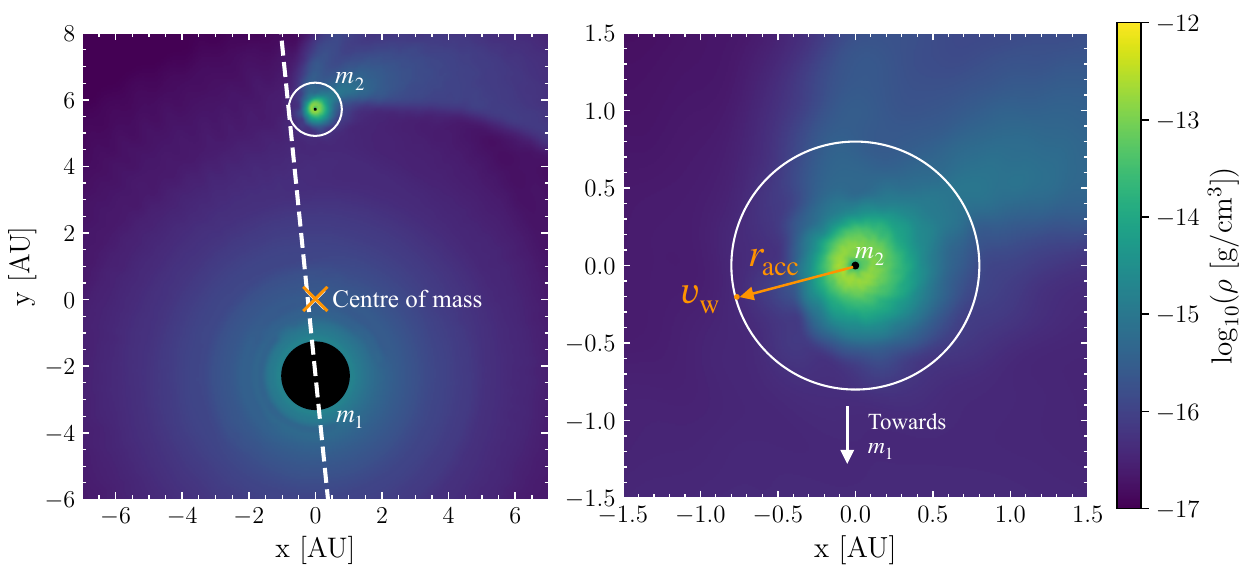}
\end{center}
\captionsetup{justification=justified}
\caption{Volume density ($\rho$) map for simulation A8 in the orbital plane. {\bf Left:} Global view showing the donor ($m_1$), the accretor ($m_2$), and the system's center of mass (cross). The dashed line is used to extract the wind velocity profiles shown in Fig.~\ref{fig:vel_real}. {\bf Right :} Close-up of the secondary and its associated accretion disk. In both panels, the circle centered on $m_2$ represents the characteristic radius $r_\mathrm{acc}$, marking the location where the upstream wind velocity ($v_\mathrm{w}$) is measured to represent the actual flow conditions encountered by the accretor.}
\label{fig:density}
\end{figure*}

An example of the density structure around a binary system is illustrated in Fig.~\ref{fig:density}, which shows the density configuration for simulation A8 in the vicinity of the binary system. For the purposes of studying the accretion process, we focus on the global accretion rates rather than the internal properties of the accretion discs.

\begin{figure}[t]
\begin{center}
\includegraphics[width=\linewidth]{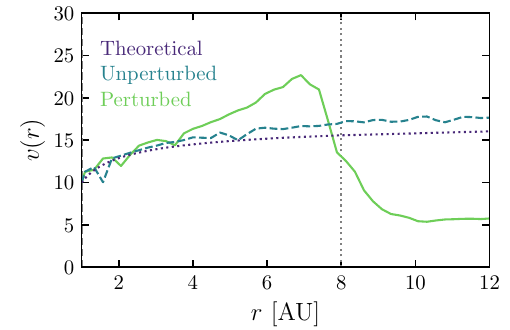}
\end{center}
\captionsetup{justification=justified} 
\caption{Velocity profiles for simulation A8, extracted extracted along the dashed axis shown in the left panel of Fig.~\ref{fig:density}. 
The perturbed profile (solid line) represents the velocity measured in the direction toward the secondary, whereas the unperturbed profile (dashed) corresponds to the opposite direction. For comparison, the theoretical wind profile ($v_{\mathrm{w,u}}$) from Fig.~\ref{fig:wind1D} is shown in a dotted line. The vertical dotted line indicates the orbital separation $a=8$ AU.}
\label{fig:vel_real}
\end{figure}

\begin{figure}[t]
\begin{center}
\includegraphics[width=\linewidth]{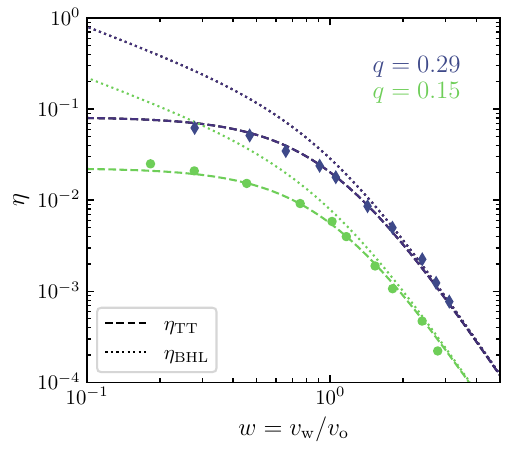}\\[-22pt]
\includegraphics[width=\linewidth]{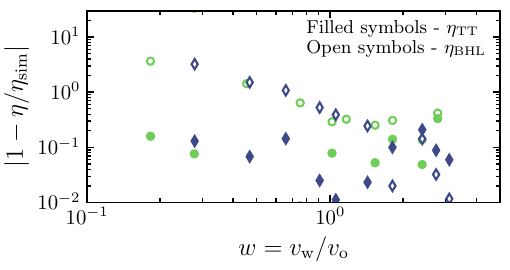}
\end{center}
\captionsetup{justification=justified}
\caption{{\bf Top:} Mass accretion efficiency $\eta$ as a function of the wind dimensionless parameter $w$. Analytical predictions from the geometrically corrected model ($\eta_\mathrm{TT}$, dashed line) and the standard BHL implementation ($\eta_\mathrm{BHL}$, dotted line) are shown for comparison. {\bf Bottom:} Fractional errors between the analytical predictions that the numerical results. In both panels, diamonds and bullets represents measurements from the \textcolor{blue}{A3--A50 ($q = 0.29$)} and \textcolor{blue}{B2.5--B40 ($q=0.15$)} simulation families, respectively. }
\label{fig:w_eta}
\end{figure}

Table~\ref{tab:set1} lists the mass accretion efficiencies obtained from our simulations ($\eta_{\mathrm{sim}}$) alongside the analytical estimates $\eta_{\mathrm{BHL}}$ and $\eta_{\mathrm{TT}}$. Calculating these analytical values requires the dimensionless parameters $q$ and $w = v_{\mathrm{w}}/v_{\mathrm{o}}$. While $q$ is fixed for each simulation set and $v_\mathrm{o}$ is constant for a given circular orbit, determining $v_{\mathrm{w}}$ is more complex, as it should correspond to the actual wind velocity encountered by the accretor. Although $v_{\mathrm{w}}$ is often assumed to match the unperturbed stellar wind velocity ($v_{\mathrm{w,u}}$) at the accretor's orbital distance (solid line in Fig.~\ref{fig:wind1D}), the binary's gravitational potential significantly alters the local flow \citep[see, e.g.,][]{Saladino2019}. As evidenced by the symbols in Fig.~\ref{fig:wind1D} and further illustrated by the velocity profile in Fig.~\ref{fig:vel_real}, the effective velocity frequently departs from the unperturbed profile. This discrepancy is most pronounced in compact orbits and slow-wind regimes, with the specific magnitude of these variations depending on the orbital separation and the wind acceleration profile.

To account for this, we directly measure the gas velocity in the vicinity of the accretor from our hydrodynamical simulations. Following the methodology illustrated in Fig.~\ref{fig:density}, we measure $v_{\mathrm{w}}$ at the leading intersection between a radial axis from the primary and a circle of radius $r_{\mathrm{acc}}$ centered on the accretor. Measuring $v_{\mathrm{w}}$ in the upstream direction is crucial to capture the actual wind velocity encountered by the accretor before the flow is further modified by the bow shock and the accretion disk. This characteristic scale is defined using the unperturbed wind velocity ($v_{\mathrm{w,u}}$) as
\begin{equation}
r_\mathrm{acc} = \frac{G m_2}{v_{\mathrm{w,u}}^{2} + v_{\mathrm{o}}^{2}}.
\label{eq:r_acc}
\end{equation}
We remark that in all simulations, $r_\mathrm{acc}$ fully encloses the accretion disc as illustrated in Fig.~\ref{fig:density}.

The results listed in Table~\ref{tab:set1} are presented in the top panel of Fig.~\ref{fig:w_eta}. Using the locally measured $v_{\mathrm{w}}$ provides a physically accurate basis for computing $w$ and comparing the analytical estimations of the mass accretion efficiencies ($\eta_\mathrm{TT}$ and $\eta_\mathrm{BHL}$) against that obtained directly from our numerical results ($\eta_\mathrm{sim}$). Our comparison shows that for simulations with $w > 1$, all three quantities converge to similar values ($\eta_\mathrm{sim} \approx \eta_\mathrm{TT} \approx \eta_\mathrm{BHL}$), although we invariably find that $\eta_\mathrm{TT} < \eta_\mathrm{BHL}$. However, in the slow-wind regime ($w < 1$), we confirm that the standard BHL prescription systematically overestimates the accretion efficiency, while the geometrically corrected model faithfully reproduces the simulation results. As quantified in the bottom panel of Fig.~\ref{fig:w_eta}, $\eta_\mathrm{TT}$  exhibits an average error of about 10\% (ranging from 1\% to 19\%), whereas $\eta_\mathrm{BHL}$ deviates by an average error of 100\%, with discrepancies escalating up to more than 400\% in certain regimes.

\section{Discussion}
\label{sec:discussion}

Table~\ref{tab:set1} and Fig.~\ref{fig:w_eta} demonstrate that the analytical predictions of \citet{TejedaToala2025} successfully replicate the overall qualitative trends and scaling behavior of the mass accretion efficiency observed in our numerical simulations. Although other empirical formalisms can also fit aspects of these data, the close alignment of our SPH results with the \citet{TejedaToala2025} framework highlights the utility of this geometric correction as a robust, parameter-free tool capable of capturing the primary dynamics of the flow even when local grid or smoothing details are unresolved.

Moreover, our analysis reinforces earlier conclusions of \citet{TejedaToala2025} that relying on an unperturbed wind profile ($v_\mathrm{w,u}$) leads to inaccurate estimates of the mass accretion rate. Instead, the appropriate velocity to employ is the local gas velocity measured in the upstream direction in the immediate vicinity of the accretor ($v_\mathrm{w}$), which faithfully represents the conditions experienced by the accreting object.

\begin{figure}[t]
\begin{center}
\includegraphics[width=\linewidth]{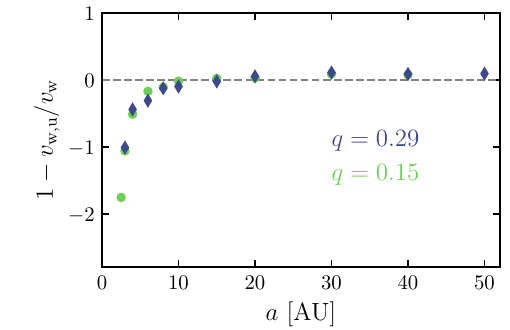}
\end{center}
\captionsetup{justification=justified}
\caption{Fractional difference between the unperturbed wind velocity ($v_{\mathrm{w,u}}$) and the local gas velocity measured from the simulations ($v_{\mathrm{w}}$), defined as $1 - v_{\mathrm{w,u}}/v_{\mathrm{w}}$.  Symbols represent results for both simulation families (A and B) as a function of orbital separation.}
\label{fig:vel_fraction}
\end{figure}

The inaccuracy of using $v_\mathrm{w,u}$ stems from the fact that the companion is orbiting; its motion induces a continuously changing gravitational field that significantly perturbs the local gas velocity field away from a standard, stationary BHL-type of flow. To illustrate this effect, Fig.~\ref{fig:vel_real} shows the velocity profile extracted along the radial axis shown in Fig.~\ref{fig:density} for simulation A8. The profile measured toward the accretor departs markedly from the expected wind solution; the gas velocity initially speeds up before sharply slowing down just before encountering the secondary. In contrast, the velocity profile extracted in the direction away from the accretor closely follows the unperturbed wind profile predicted by {\sc phantom}.

To further investigate these discrepancies, Fig.~\ref{fig:vel_fraction} presents the fractional difference between the local gas velocity experienced by the accretor and the value expected from the unperturbed wind profile, defined as $1 - v_\mathrm{w,u}/v_\mathrm{w}$. The largest deviations occur in simulations with orbital separations $a \leq 10$ AU, which correspond to the region where most of the wind acceleration takes place in our {\sc phantom} models.

\begin{figure}[t!]
\begin{center}
\includegraphics[width=\linewidth]{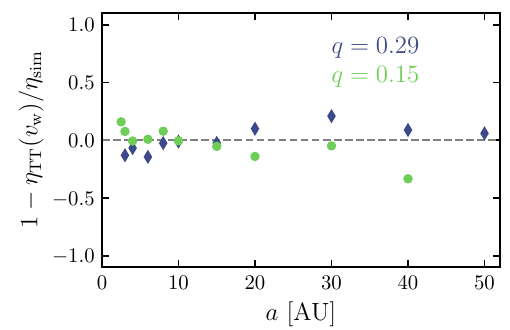}\\[-19pt]
\includegraphics[width=\linewidth]{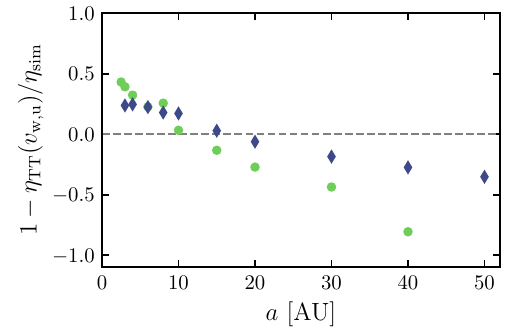}\\[-19pt]
\includegraphics[width=\linewidth]{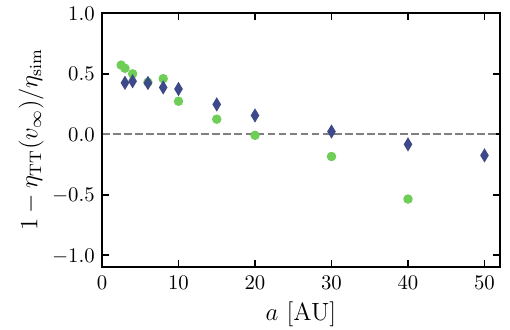}
\end{center}
\captionsetup{justification=justified}
\caption{Accuracy of the $\eta_{\mathrm{TT}}$ analytical prescription using different velocity estimates relative to the simulated efficiency ($\eta_{\mathrm{sim}}$). The panels show the fractional error ($1 - \eta_{\mathrm{TT}}/\eta_{\mathrm{sim}}$) using: the local upstream velocity ($v_{\mathrm{w}}$, top), the unperturbed wind velocity ($v_{\mathrm{w,u}}$, middle), and the terminal wind velocity ($v_{\infty}$, bottom). Symbols represent results for both simulation families as a function of orbital separation.}
\label{fig:eta_fraction}
\end{figure}

A complementary analysis is shown in Fig.~\ref{fig:eta_fraction}, where we evaluate the accuracy of the $\eta_\mathrm{TT}$ prescription using three different velocity inputs: the upstream local velocity $v_\mathrm{w}$ (top panel), the unperturbed wind velocity $v_\mathrm{w,u}$ (middle panel), and the terminal wind velocity $v_\infty$ (bottom panel). The top panel provides an alternative representation of Fig.~\ref{fig:w_eta}-bottom, confirming that adopting the local wind velocity near the accretor reproduces the simulated accretion efficiencies with high fidelity. In contrast, relying on the unperturbed wind velocity (middle panel) introduces errors of $\sim50\%$ for compact orbits and up to $\sim80\%$ for the widest configurations in Family B. Nevertheless, these discrepancies remain significantly smaller than those produced by the standard BHL formulation.

In practice, observers rarely have access to the perturbed local flow near the accretor, relying instead on estimates of the donor’s terminal wind velocity, $v_\infty$. The bottom panel of Fig.~\ref{fig:eta_fraction} shows that combining $v_\infty$ with the $\eta_\mathrm{TT}$ model yields acceptable uncertainties, typically within $\sim 50\%$ for the inferred mass accretion efficiency. This indicates that the purely geometric framework offers a reliable approximation of the global accretion behavior, capturing the essential scaling trends even when complex local flow details remain unresolved.

\subsection{Comparison with previous works}

We acknowledge the substantial body of high-quality research addressing accretion in binary systems within the slow-wind regime \citep[e.g.,][and references therein]{Theuns1996,Nagae2004,deValBorro2009,Liu2017,Saladino2019,Lee2022,Malfait2024}. Typically, these studies compare numerically-derived accretion efficiencies against predictions from the standard BHL formulation, consistently reporting significant discrepancies. We identify two primary reasons for these departures.

First, many adapted BHL prescriptions introduce an ad hoc efficiency parameter, $\alpha_\mathrm{BHL}$, to reconcile simulations with analytical expectations (see Section~\ref{sec:intro}). In contrast, the $\eta_\mathrm{TT}$ model accounts for the inherent tilt of the accretion cylinder caused by the relative orbital motion of the binary. This geometric correction naturally reduces the effective accretion cross-section, allowing the model to recover simulated results across all wind regimes without empirical tuning.

Second, several studies compute mass accretion efficiencies using wind velocities derived from unperturbed, single-star wind profiles. As demonstrated in our analysis, this approach neglects the strong dynamical influence of the accretor, which perturbs the local wind velocity experienced by the accreting object. This effect is most pronounced at the small orbital separations explored here ($a < 10$~AU).

\subsection{On Symbiotic Systems}
\label{sec:SySts}

Improving our understanding of the wind accretion mechanism in the slow-wind regime has a direct impact on the interpretation of observed symbiotic systems, where a WD accretes from the wind of a RG or AGB star. These late-type stars typically exhibit stellar wind velocities of $v_\mathrm{w}$=5--20 km~s$^{-1}$ \citep[e.g.,][]{Ramstedt2020,Wallstrom2025} and orbital velocities of $v_\mathrm{o}$ = 30--50~km~s$^{-1}$. Consequently, symbiotic systems have typical $w$ values below 1.

\begin{figure}
\begin{center}
\includegraphics[width=\linewidth]{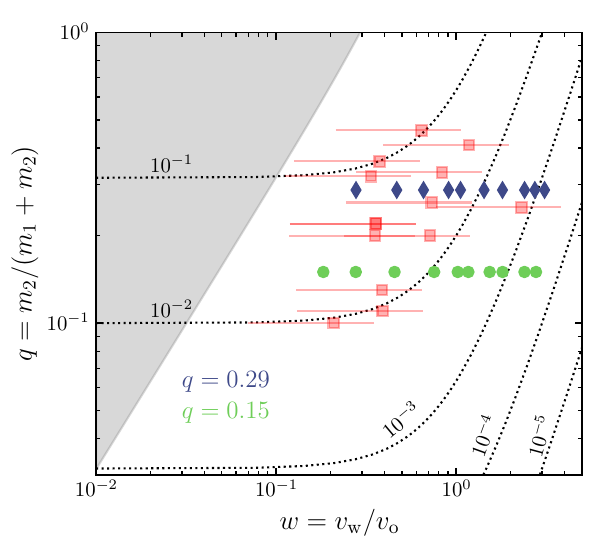}
\end{center}
\captionsetup{justification=justified}
\caption{Parameter space of mass accretion efficiency in the $q$--$w$ plane. Symbols represent the two simulation families (A and B), while dashed lines indicate contours of constant $\log_{10}(\eta_{\mathrm{TT}})$. The square symbols with error bars denote the properties of observed symbiotic systems (see Table~\ref{tab:objects}). The shaded region at high $q$ and low $w$ represents the portion of the parameter space where the analytical model of \citet{TejedaToala2025} is formally invalid due to the breakdown of the underlying geometric assumptions.}
\label{fig:eta_w}
\end{figure}

The SPH simulations presented here are intended to broadly encompass the observed properties of known symbiotic systems. For example, Fig.~\ref{fig:eta_w} illustrates the parameter space occupied by the two families of simulations within a $q-w$ diagram. The figure includes contours of constant log$_{10}(\eta)$ values alongside the positions of observed symbiotic systems (see Appendix~A for details on the data sources). In general, the observations and simulations occupy a similar region in the $q-w$ diagram, spanning efficiencies between the log$_{10}(\eta) = -4$ and $-$1. 

The accretion luminosity in a symbiotic system can be estimated as
\begin{equation}
    L_\mathrm{acc} = \frac{1}{2} \frac{G m_\mathrm{WD}}{R_\mathrm{WD}} \dot{M}_\mathrm{acc},
\end{equation}
\noindent where $m_\mathrm{WD}$ and $R_\mathrm{WD}$ are the mass and radius of the accreting WD, respectively, and $\dot{M}_\mathrm{acc} = \eta_\mathrm{sim} \dot{M}_\mathrm{w}$ is the mass accretion rate. Adopting a typical radius of $R_\mathrm{WD}=0.01$~R$_\odot$ \citep[e.g.,][]{Boshkayev2016,Pasquini2023,Karinkuzhi2024} and using the accretor masses from our simulations, we estimate accretion luminosities in the range of $L_\mathrm{acc}=[0.04 - 23] \times10^{33}$~erg~s$^{-1} = [0.01 - 6]$~L$_\odot$. These values are consistent with the X-ray luminosities reported for symbiotic systems \citep[see, e.g.,][and references therein]{Guerrero2024}.

\section{Summary}
\label{sec:summary}

In this work, we have presented a suite of SPH simulations designed to test wind accretion in binary systems operating in the slow-wind regime ($w < 1$). Our primary goal was to assess the validity of analytical prescriptions for the mass accretion efficiency by directly comparing numerical results with both the standard implementation of the BHL formalism and the geometrically corrected model proposed by \citet{TejedaToala2025}.

We explored accretion in circular binary configurations and demonstrated that the standard BHL implementation systematically overestimates the mass accretion efficiency in the slow-wind regime ($w <1$). In contrast, the geometrically-corrected model reproduces the accretion efficiencies measured directly from the simulations remarkably well across the full range of explored parameters. A fundamental aspect of this agreement is the geometric foundation of the model: as the accretor orbits the donor star, the accretion cylinder is not aligned with the radial wind flow but is instead tilted by the relative orbital motion. This wind aberration effectively reduces the accretion cross-section, a physical correction that is naturally captured by the $\eta_\mathrm{TT}$ prescription. Consequently, the model recovers the measured efficiencies across all wind regimes without the need for empirical efficiency factors often required in traditional BHL implementations.

As is well established in binary wind accretion studies, the relevant wind velocity for accretion estimates is not the value derived from the unperturbed wind profile ($v_{\mathrm{w,u}}$), but rather the local gas velocity measured upstream in the vicinity of the accretor ($v_{\mathrm{w}}$). Our simulations confirm that the gravitational potential of the accretor significantly perturbs the wind flow, modifying the effective relative velocity. This effect is especially pronounced within the wind acceleration zone of the primary star. While using the locally measured velocity is essential for achieving formal agreement between analytical predictions and numerical results. We note that for observed systems, adopting the terminal wind velocity typically yields errors of $\lesssim 50\%$ with the $\eta_\mathrm{TT}$ prescription. Crucially, this discrepancy remains significantly smaller than the order-of-magnitude errors produced by the standard BHL implementation in the slow-wind regime.

Furthermore, our numerical results presented here provide a robust benchmark for the wind accretion approximation developed by \citet{TejedaToala2025}. As this prescription has already been applied by \citet{Maldonado2025,Maldonado2025_WRLO} to study the long-term evolution of accreting white dwarfs in binary systems, our work provides the necessary hydrodynamical validation for those studies.

Overall, our results demonstrate that the geometrically corrected wind accretion model successfully captures the primary scaling trends and qualitative behavior of the flow, offering a reliable analytical description of the mass accretion efficiency in the slow-wind regime without the need for adjustable parameters. This provides a physically motivated framework for modeling wind-fed accretion in a wide variety of binary systems, ranging from symbiotic stars to other interacting binaries with evolved donors.

%----------------------------------------------------------

\renewcommand{\refname}{REFERENCES}
\bibliography{rmaa}

@ARTICLE{VasquezTorresRAquarii,
       author = {{Vasquez-Torres}, D.~A. and {Toal{\'a}}, J.~A. and {Sacchi}, A. and {Guerrero}, M.~A. and {Tejeda}, E. and {Karovska}, M. and {Montez}, R., Jr.},
        title = "{The impact of periastron passage on the X-ray and optical properties of the Symbiotic System R Aquarii}",
      journal = {\mnras},
         year = 2024,
        month = dec,
       volume = {535},
       number = {3},
        pages = {2724-2741},
          doi = {10.1093/mnras/stae2538},
archivePrefix = {arXiv},
       eprint = {2411.04935},
 primaryClass = {astro-ph.SR},
       adsurl = {https://ui.adsabs.harvard.edu/abs/2024MNRAS.535.2724V}
}

@BOOK{Spizer1978,
       author = {{Spitzer}, Lyman},
        title = "{Physical processes in the interstellar medium}",
         year = 1978,
          doi = {10.1002/9783527617722},
       adsurl = {https://ui.adsabs.harvard.edu/abs/1978ppim.book.....S}
}

@ARTICLE{Pod2007,
       author = {{Podsiadlowski}, Ph. and {Mohamed}, S.},
        title = "{The Origin and Evolution of Symbiotic Binaries}",
      journal = {Baltic Astronomy},
         year = 2007,
        month = jan,
       volume = {16},
        pages = {26-33},
       adsurl = {https://ui.adsabs.harvard.edu/abs/2007BaltA..16...26P}
}

@ARTICLE{Hainich2020,
       author = {{Hainich}, R. and {Oskinova}, L.~M. and {Torrej{\'o}n}, J.~M. and {Fuerst}, F. and {Bodaghee}, A. and {Shenar}, T. and {Sander}, A.~A.~C. and {Todt}, H. and {Spetzer}, K. and {Hamann}, W.-R.},
        title = "{The stellar and wind parameters of six prototypical HMXBs and their evolutionary status}",
      journal = {\aap},
         year = 2020,
        month = feb,
       volume = {634},
          eid = {A49},
        pages = {A49},
          doi = {10.1051/0004-6361/201935498},
archivePrefix = {arXiv},
       eprint = {2001.02420},
 primaryClass = {astro-ph.SR},
       adsurl = {https://ui.adsabs.harvard.edu/abs/2020A&A...634A..49H}
}

@ARTICLE{Mohamed2012,
       author = {{Mohamed}, S. and {Podsiadlowski}, Ph.},
        title = "{Mass Transfer in Mira-type Binaries}",
      journal = {Baltic Astronomy},
         year = 2012,
        month = jan,
       volume = {21},
        pages = {88-96},
          doi = {10.1515/astro-2017-0362},
       adsurl = {https://ui.adsabs.harvard.edu/abs/2012BaltA..21...88M}
}

@INPROCEEDINGS{Mohamed2007,
       author = {{Mohamed}, S. and {Podsiadlowski}, Ph.},
        title = "{Wind Roche-Lobe Overflow: a New Mass-Transfer Mode for Wide Binaries}",
    booktitle = {15th European Workshop on White Dwarfs},
         year = 2007,
       editor = {{Napiwotzki}, R. and {Burleigh}, M.~R.},
       series = {Astronomical Society of the Pacific Conference Series},
       volume = {372},
        month = sep,
        pages = {397},
       adsurl = {https://ui.adsabs.harvard.edu/abs/2007ASPC..372..397M}
}

@ARTICLE{Siess2022,
       author = {{Siess}, L. and {Homan}, W. and {Toupin}, S. and {Price}, D.~J.},
        title = "{3D simulations of AGB stellar winds. I. Steady winds and dust formation}",
      journal = {\aap},
         year = 2022,
        month = nov,
       volume = {667},
          eid = {A75},
        pages = {A75},
          doi = {10.1051/0004-6361/202243540},
archivePrefix = {arXiv},
       eprint = {2208.13869},
 primaryClass = {astro-ph.SR},
       adsurl = {https://ui.adsabs.harvard.edu/abs/2022A&A...667A..75S}
}

@ARTICLE{HuarteEspinosa2013,
       author = {{Huarte-Espinosa}, M. and {Carroll-Nellenback}, J. and {Nordhaus}, J. and {Frank}, A. and {Blackman}, E.~G.},
        title = "{The formation and evolution of wind-capture discs in binary systems}",
      journal = {\mnras},
         year = 2013,
        month = jul,
       volume = {433},
       number = {1},
        pages = {295-306},
          doi = {10.1093/mnras/stt725},
archivePrefix = {arXiv},
       eprint = {1211.1672},
 primaryClass = {astro-ph.SR},
       adsurl = {https://ui.adsabs.harvard.edu/abs/2013MNRAS.433..295H}
}

@ARTICLE{McDonald2018,
       author = {{McDonald}, I. and {De Beck}, E. and {Zijlstra}, A.~A. and {Lagadec}, E.},
        title = "{Pulsation-triggered dust production by asymptotic giant branch stars}",
      journal = {\mnras},
         year = 2018,
        month = dec,
       volume = {481},
       number = {4},
        pages = {4984-4999},
          doi = {10.1093/mnras/sty2607},
archivePrefix = {arXiv},
       eprint = {1809.07965},
 primaryClass = {astro-ph.SR},
       adsurl = {https://ui.adsabs.harvard.edu/abs/2018MNRAS.481.4984M}
}

@ARTICLE{Vassiliadis1993,
       author = {{Vassiliadis}, E. and {Wood}, P.~R.},
        title = "{Evolution of Low- and Intermediate-Mass Stars to the End of the Asymptotic Giant Branch with Mass Loss}",
      journal = {\apj},
         year = 1993,
        month = aug,
       volume = {413},
        pages = {641},
          doi = {10.1086/173033},
       adsurl = {https://ui.adsabs.harvard.edu/abs/1993ApJ...413..641V}
}

@ARTICLE{Esseldeurs2023,
       author = {{Esseldeurs}, M. and {Siess}, L. and {De Ceuster}, F. and {Homan}, W. and {Malfait}, J. and {Maes}, S. and {Konings}, T. and {Ceulemans}, T. and {Decin}, L.},
        title = "{3D simulations of AGB stellar winds. II. Ray-tracer implementation and impact of radiation on the outflow morphology}",
      journal = {\aap},
         year = 2023,
        month = jun,
       volume = {674},
          eid = {A122},
        pages = {A122},
          doi = {10.1051/0004-6361/202346282},
archivePrefix = {arXiv},
       eprint = {2304.09786},
 primaryClass = {astro-ph.SR},
       adsurl = {https://ui.adsabs.harvard.edu/abs/2023A&A...674A.122E}
}

@ARTICLE{Price2018,
       author = {{Price}, Daniel J. and {Wurster}, James and {Tricco}, Terrence S. and {Nixon}, Chris and {Toupin}, St{\'e}ven and {Pettitt}, Alex and {Chan}, Conrad and {Mentiplay}, Daniel and {Laibe}, Guillaume and {Glover}, Simon and {Dobbs}, Clare and {Nealon}, Rebecca and {Liptai}, David and {Worpel}, Hauke and {Bonnerot}, Cl{\'e}ment and {Dipierro}, Giovanni and {Ballabio}, Giulia and {Ragusa}, Enrico and {Federrath}, Christoph and {Iaconi}, Roberto and {Reichardt}, Thomas and {Forgan}, Duncan and {Hutchison}, Mark and {Constantino}, Thomas and {Ayliffe}, Ben and {Hirsh}, Kieran and {Lodato}, Giuseppe},
        title = "{Phantom: A Smoothed Particle Hydrodynamics and Magnetohydrodynamics Code for Astrophysics}",
      journal = {PASA},
         year = 2018,
        month = sep,
       volume = {35},
          eid = {e031},
        pages = {e031},
          doi = {10.1017/pasa.2018.25},
archivePrefix = {arXiv},
       eprint = {1702.03930},
 primaryClass = {astro-ph.IM},
       adsurl = {https://ui.adsabs.harvard.edu/abs/2018PASA...35...31P}
}

@ARTICLE{Davidson1973,
       author = {{Davidson}, Kris and {Ostriker}, Jeremiah P.},
        title = "{Neutron-Star Accretion in a Stellar Wind: Model for a Pulsed X-Ray Source}",
      journal = {\apj},
         year = 1973,
        month = jan,
       volume = {179},
        pages = {585-598},
          doi = {10.1086/151897},
       adsurl = {https://ui.adsabs.harvard.edu/abs/1973ApJ...179..585D}
}

@ARTICLE{Liu2017,
       author = {{Liu}, Zheng-Wei and {Stancliffe}, Richard J. and {Abate}, Carlo and {Matrozis}, Elvijs},
        title = "{Three-dimensional Hydrodynamical Simulations of Mass Transfer in Binary Systems by a Free Wind}",
      journal = {\apj},
         year = 2017,
        month = sep,
       volume = {846},
       number = {2},
          eid = {117},
        pages = {117},
          doi = {10.3847/1538-4357/aa8622},
archivePrefix = {arXiv},
       eprint = {1708.03639},
 primaryClass = {astro-ph.SR},
       adsurl = {https://ui.adsabs.harvard.edu/abs/2017ApJ...846..117L}
}

@ARTICLE{Theuns1996,
       author = {{Theuns}, Tom and {Boffin}, Henri M.~J. and {Jorissen}, Alain},
        title = "{Wind accretion in binary stars - II. Accretion rates}",
      journal = {\mnras},
         year = 1996,
        month = jun,
       volume = {280},
       number = {4},
        pages = {1264-1276},
          doi = {10.1093/mnras/280.4.1264},
archivePrefix = {arXiv},
       eprint = {astro-ph/9602089},
 primaryClass = {astro-ph},
       adsurl = {https://ui.adsabs.harvard.edu/abs/1996MNRAS.280.1264T}
}

@INCOLLECTION{Millar2004,
       author = {{Millar}, Tom J.},
        title = "{Molecule and Dust Grain Formation}",
    booktitle = {Asymptotic Giant Branch Stars},
         year = 2004,
       editor = {{Habing}, Harm J. and {Olofsson}, Hans},
        pages = {247-289},
          doi = {10.1007/978-1-4757-3876-6_5},
       adsurl = {https://ui.adsabs.harvard.edu/abs/2004agbs.book..247M}
}

@ARTICLE{Saladino2019,
       author = {{Saladino}, M.~I. and {Pols}, O.~R. and {Abate}, C.},
        title = "{Slowly, slowly in the wind. 3D hydrodynamical simulations of wind mass transfer and angular-momentum loss in AGB binary systems}",
      journal = {\aap},
         year = 2019,
        month = jun,
       volume = {626},
          eid = {A68},
        pages = {A68},
          doi = {10.1051/0004-6361/201834598},
archivePrefix = {arXiv},
       eprint = {1903.04515},
 primaryClass = {astro-ph.SR},
       adsurl = {https://ui.adsabs.harvard.edu/abs/2019A&A...626A..68S}
}

@ARTICLE{SaladinoPols2019,
       author = {{Saladino}, M.~I. and {Pols}, O.~R.},
        title = "{The eccentric behaviour of windy binary stars}",
      journal = {\aap},
         year = 2019,
        month = sep,
       volume = {629},
          eid = {A103},
        pages = {A103},
          doi = {10.1051/0004-6361/201935625},
archivePrefix = {arXiv},
       eprint = {1906.02038},
 primaryClass = {astro-ph.SR},
       adsurl = {https://ui.adsabs.harvard.edu/abs/2019A&A...629A.103S}
}

@ARTICLE{Lee2022,
       author = {{Lee}, Young-Min and {Kim}, Hyosun and {Lee}, Hee-Won},
        title = "{Formation of the Asymmetric Accretion Disk from Stellar Wind Accretion in an S-type Symbiotic Star}",
      journal = {\apj},
         year = 2022,
        month = jun,
       volume = {931},
       number = {2},
          eid = {142},
        pages = {142},
          doi = {10.3847/1538-4357/ac67d6},
archivePrefix = {arXiv},
       eprint = {2205.04758},
 primaryClass = {astro-ph.SR},
       adsurl = {https://ui.adsabs.harvard.edu/abs/2022ApJ...931..142L}
}

@ARTICLE{Malfait2024,
       author = {{Malfait}, J. and {Siess}, L. and {Esseldeurs}, M. and {De Ceuster}, F. and {Wallstr{\"o}m}, S.~H.~J. and {de Koter}, A. and {Decin}, L.},
        title = "{Impact of H I cooling and study of accretion disks in asymptotic giant branch wind-companion smoothed particle hydrodynamic simulations}",
      journal = {\aap},
         year = 2024,
        month = nov,
       volume = {691},
          eid = {A84},
        pages = {A84},
          doi = {10.1051/0004-6361/202450338},
archivePrefix = {arXiv},
       eprint = {2408.13158},
 primaryClass = {astro-ph.SR},
       adsurl = {https://ui.adsabs.harvard.edu/abs/2024A&A...691A..84M}
}

@INPROCEEDINGS{Boffin2015,
       author = {{Boffin}, Henri M.~J.},
        title = "{Mass Transfer by Stellar Wind}",
    booktitle = {Astrophysics and Space Science Library},
         year = 2015,
       editor = {{Boffin}, Henri M.~J. and {Carraro}, Giovanni and {Beccari}, Giacomo},
       series = {Astrophysics and Space Science Library},
       volume = {413},
        month = jan,
        pages = {153},
          doi = {10.1007/978-3-662-44434-4_7},
archivePrefix = {arXiv},
       eprint = {1406.3473},
 primaryClass = {astro-ph.SR},
       adsurl = {https://ui.adsabs.harvard.edu/abs/2015ASSL..413..153B}
}

@ARTICLE{Nagae2004,
       author = {{Nagae}, T. and {Oka}, K. and {Matsuda}, T. and {Fujiwara}, H. and {Hachisu}, I. and {Boffin}, H.~M.~J.},
        title = "{Wind accretion in binary stars. I. Mass accretion ratio}",
      journal = {\aap},
         year = 2004,
        month = may,
       volume = {419},
        pages = {335-343},
          doi = {10.1051/0004-6361:20040070},
archivePrefix = {arXiv},
       eprint = {astro-ph/0403329},
 primaryClass = {astro-ph},
       adsurl = {https://ui.adsabs.harvard.edu/abs/2004A&A...419..335N}
}

@ARTICLE{Weng2024,
       author = {{Weng}, Shan-Shan and {Ji}, Long},
        title = "{X-Ray Views of Galactic Accreting Pulsars in High-Mass X-Ray Binaries}",
      journal = {Universe},
         year = 2024,
        month = dec,
       volume = {10},
       number = {12},
          eid = {453},
        pages = {453},
          doi = {10.3390/universe10120453},
archivePrefix = {arXiv},
       eprint = {2412.17275},
 primaryClass = {astro-ph.HE},
       adsurl = {https://ui.adsabs.harvard.edu/abs/2024Univ...10..453W}
}

@ARTICLE{Shakura2015,
       author = {{Shakura}, N.~I. and {Postnov}, K.~A. and {Kochetkova}, A. Yu. and {Hjalmarsdotter}, L. and {Sidoli}, L. and {Paizis}, A.},
        title = "{Wind accretion: Theory and observations}",
      journal = {Astronomy Reports},
         year = 2015,
        month = jul,
       volume = {59},
       number = {7},
        pages = {645-655},
          doi = {10.1134/S1063772915070112},
archivePrefix = {arXiv},
       eprint = {1407.3163},
 primaryClass = {astro-ph.HE},
       adsurl = {https://ui.adsabs.harvard.edu/abs/2015ARep...59..645S}
}

@ARTICLE{Bondi1944,
       author = {{Bondi}, H. and {Hoyle}, F.},
        title = "{On the mechanism of accretion by stars}",
      journal = {\mnras},
         year = 1944,
        month = jan,
       volume = {104},
        pages = {273},
          doi = {10.1093/mnras/104.5.273},
       adsurl = {https://ui.adsabs.harvard.edu/abs/1944MNRAS.104..273B}
}

@ARTICLE{HoyleLyttleton1939,
       author = {{Hoyle}, F. and {Lyttleton}, R.~A.},
        title = "{The effect of interstellar matter on climatic variation}",
      journal = {Proceedings of the Cambridge Philosophical Society},
         year = 1939,
        month = jan,
       volume = {35},
       number = {3},
        pages = {405},
          doi = {10.1017/S0305004100021150},
       adsurl = {https://ui.adsabs.harvard.edu/abs/1939PCPS...35..405H}
}

@ARTICLE{TejedaToala2025,
       author = {{Tejeda}, Emilio and {Toal{\'a}}, Jes{\'u}s A.},
        title = "{Geometric Correction for Wind Accretion in Binary Systems}",
      journal = {\apj},
         year = 2025,
        month = feb,
       volume = {980},
       number = {2},
          eid = {226},
        pages = {226},
          doi = {10.3847/1538-4357/ada953},
archivePrefix = {arXiv},
       eprint = {2411.01755},
 primaryClass = {astro-ph.HE},
       adsurl = {https://ui.adsabs.harvard.edu/abs/2025ApJ...980..226T}
}

@INPROCEEDINGS{Boshkayev2016,
       author = {{Boshkayev}, K.~A. and {Rueda}, J.~A. and {Zhami}, B.~A. and {Kalymova}, Zh. A. and {Balgymbekov}, G. Sh.},
        title = "{Equilibrium structure of white dwarfs at finite temperatures}",
    booktitle = {International Journal of Modern Physics Conference Series},
         year = 2016,
       series = {International Journal of Modern Physics Conference Series},
       volume = {41},
        month = mar,
          eid = {1660129},
        pages = {1660129},
          doi = {10.1142/S2010194516601290},
archivePrefix = {arXiv},
       eprint = {1510.02024},
 primaryClass = {astro-ph.SR},
       adsurl = {https://ui.adsabs.harvard.edu/abs/2016IJMPS..4160129B}
}

@ARTICLE{deValBorro2009,
       author = {{de Val-Borro}, M. and {Karovska}, M. and {Sasselov}, D.},
        title = "{Numerical Simulations of Wind Accretion in Symbiotic Binaries}",
      journal = {\apj},
         year = 2009,
        month = aug,
       volume = {700},
       number = {2},
        pages = {1148-1160},
          doi = {10.1088/0004-637X/700/2/1148},
archivePrefix = {arXiv},
       eprint = {0905.3542},
 primaryClass = {astro-ph.SR},
       adsurl = {https://ui.adsabs.harvard.edu/abs/2009ApJ...700.1148D}
}

@INPROCEEDINGS{Mikolajewska2003,
       author = {{Miko{\l}ajewska}, J.},
        title = "{Orbital and Stellar Parameters of Symbiotic Stars (invited review talks)}",
    booktitle = {Symbiotic Stars Probing Stellar Evolution},
         year = 2003,
       editor = {{Corradi}, R.~L.~M. and {Mikolajewska}, J. and {Mahoney}, T.~J.},
       series = {Astronomical Society of the Pacific Conference Series},
       volume = {303},
        month = jan,
        pages = {9},
          doi = {10.48550/arXiv.astro-ph/0210489},
archivePrefix = {arXiv},
       eprint = {astro-ph/0210489},
 primaryClass = {astro-ph},
       adsurl = {https://ui.adsabs.harvard.edu/abs/2003ASPC..303....9M}
}

@ARTICLE{Merc2025,
       author = {{Merc}, Jaroslav},
        title = "{Symbiotic Stars in the Era of Modern Ground- and Space-Based Surveys}",
      journal = {Galaxies},
         year = 2025,
        month = apr,
       volume = {13},
       number = {3},
          eid = {49},
        pages = {49},
          doi = {10.3390/galaxies13030049},
archivePrefix = {arXiv},
       eprint = {2504.16825},
 primaryClass = {astro-ph.SR},
       adsurl = {https://ui.adsabs.harvard.edu/abs/2025Galax..13...49M}
}

@ARTICLE{Guerrero2024,
       author = {{Guerrero}, M.~A. and {Montez}, R. and {Ortiz}, R. and {Toal{\'a}}, J.~A. and {Kastner}, J.~H.},
        title = "{Asymptotic giant branch stars in the eROSITA-DE eRASS1 catalog}",
      journal = {\aap},
         year = 2024,
        month = sep,
       volume = {689},
          eid = {A62},
        pages = {A62},
          doi = {10.1051/0004-6361/202450155},
archivePrefix = {arXiv},
       eprint = {2407.10552},
 primaryClass = {astro-ph.SR},
       adsurl = {https://ui.adsabs.harvard.edu/abs/2024A&A...689A..62G}
}

@ARTICLE{Karinkuzhi2024,
       author = {{Karinkuzhi}, Drisya and {Mukhopadhyay}, Banibrata and {Wickramasinghe}, Dayal and {Tout}, Christopher A.},
        title = "{Mass-radius relation for magnetized white dwarfs from SDSS}",
      journal = {\mnras},
         year = 2024,
        month = apr,
       volume = {529},
       number = {4},
        pages = {4577-4584},
          doi = {10.1093/mnras/stae829},
archivePrefix = {arXiv},
       eprint = {2403.13888},
 primaryClass = {astro-ph.SR},
       adsurl = {https://ui.adsabs.harvard.edu/abs/2024MNRAS.529.4577K}
}

@ARTICLE{Maldonado2025,
       author = {{Maldonado}, Raul F. and {Toal{\'a}}, Jes{\'u}s A. and {Rodr{\'\i}guez-Gonz{\'a}lez}, Janis B. and {Tejeda}, Emilio},
        title = "{The Impact of Wind Accretion in Evolving Symbiotic Systems}",
      journal = {\apj},
         year = 2025,
        month = aug,
       volume = {989},
       number = {1},
          eid = {108},
        pages = {108},
          doi = {10.3847/1538-4357/ade9a5},
archivePrefix = {arXiv},
       eprint = {2502.11325},
 primaryClass = {astro-ph.SR},
       adsurl = {https://ui.adsabs.harvard.edu/abs/2025ApJ...989..108M}
}

@ARTICLE{Maldonado2025_WRLO,
       author = {{Maldonado}, Ra{\'u}l F. and {Toal{\'a}}, Jes{\'u}s A. and {Tejeda}, Emilio and {Rodr{\'\i}guez-Gonz{\'a}lez}, Janis B.},
        title = "{Entering the Wind Roche Lobe Overflow realm in Symbiotic Systems}",
      journal = {\mnras},
         year = 2025,
        month = dec,
       volume = {544},
       number = {2},
        pages = {2387-2400},
          doi = {10.1093/mnras/staf1773},
archivePrefix = {arXiv},
       eprint = {2508.06727},
 primaryClass = {astro-ph.SR},
       adsurl = {https://ui.adsabs.harvard.edu/abs/2025MNRAS.544.2387M}
}

@ARTICLE{Merc2019,
       author = {{Merc}, Jaroslav and {G{\'a}lis}, Rudolf and {Wolf}, Marek},
        title = "{First Release of the New Online Database of Symbiotic Variables}",
      journal = {Research Notes of the American Astronomical Society},
         year = 2019,
        month = feb,
       volume = {3},
       number = {2},
          eid = {28},
        pages = {28},
          doi = {10.3847/2515-5172/ab0429},
       adsurl = {https://ui.adsabs.harvard.edu/abs/2019RNAAS...3...28M}
}

@INPROCEEDINGS{Alcolea2023,
       author = {{Alcolea}, J. and {Mikolajewska}, J. and {G{\'o}mez-Garrido}, M. and {Bujarrabal}, V. and {Ilkiewicz}, K. and {Castro-Carrizo}, A. and {Desmurs}, J.-F. and {Santander-Garc{\'\i}a}, M. and {Bachiller}, R.},
        title = "{Determining the orbital parameters of binary systems with an AGB primary. The case of the R Aqr symbiotic system}",
    booktitle = {Highlights on Spanish Astrophysics XI},
         year = 2023,
       editor = {{Manteiga}, M. and {Bellot}, L. and {Benavidez}, P. and {de Lorenzo-C{\'a}ceres}, A. and {Fuente}, M.~A. and {Mart{\'\i}nez}, M.~J. and {V{\'a}zquez Acosta}, M. and {Dafonte}, C.},
        month = may,
        pages = {190},
       adsurl = {https://ui.adsabs.harvard.edu/abs/2023hsa..conf..190A}
}

@ARTICLE{PadillaLopez2026,
       author = {{Padilla-L{\'o}pez}, P. and {Maldonado}, R.~F. and {Toal{\'a}}, J.~A. and {Tejeda}, E. and {Rodr{\'\i}quez-Gonz{\'a}lez}, J.~B.},
        title = "{Wind accretion onto planets orbiting an evolving Solar-like star and their detectability}",
      journal = {arXiv e-prints},
         year = 2026,
        month = mar,
          eid = {arXiv:2603.18525},
        pages = {arXiv:2603.18525},
          doi = {10.48550/arXiv.2603.18525},
archivePrefix = {arXiv},
       eprint = {2603.18525},
 primaryClass = {astro-ph.EP},
       adsurl = {https://ui.adsabs.harvard.edu/abs/2026arXiv260318525P}
}

@ARTICLE{Pasquini2023,
       author = {{Pasquini}, L. and {Pala}, A.~F. and {Salaris}, M. and {Ludwig}, H.-G. and {Le{\~a}o}, I. and {Weiss}, A. and {de Medeiros}, J.~R.},
        title = "{Accurate mass-radius ratios for Hyades white dwarfs}",
      journal = {\mnras},
         year = 2023,
        month = jul,
       volume = {522},
       number = {3},
        pages = {3710-3718},
          doi = {10.1093/mnras/stad1252},
archivePrefix = {arXiv},
       eprint = {2304.10485},
 primaryClass = {astro-ph.SR},
       adsurl = {https://ui.adsabs.harvard.edu/abs/2023MNRAS.522.3710P}
}

@ARTICLE{Ramstedt2020,
       author = {{Ramstedt}, S. and {Vlemmings}, W.~H.~T. and {Doan}, L. and {Danilovich}, T. and {Lindqvist}, M. and {Saberi}, M. and {Olofsson}, H. and {De Beck}, E. and {Groenewegen}, M.~A.~T. and {H{\"o}fner}, S. and {Kastner}, J.~H. and {Kerschbaum}, F. and {Khouri}, T. and {Maercker}, M. and {Montez}, R. and {Quintana-Lacaci}, G. and {Sahai}, R. and {Tafoya}, D. and {Zijlstra}, A.},
        title = "{DEATHSTAR: Nearby AGB stars with the Atacama Compact Array. I. CO envelope sizes and asymmetries: A new hope for accurate mass-loss-rate estimates}",
      journal = {\aap},
         year = 2020,
        month = aug,
       volume = {640},
          eid = {A133},
        pages = {A133},
          doi = {10.1051/0004-6361/201936874},
archivePrefix = {arXiv},
       eprint = {2008.07885},
 primaryClass = {astro-ph.SR},
       adsurl = {https://ui.adsabs.harvard.edu/abs/2020A&A...640A.133R}
}

@ARTICLE{Vathachira2026,
       author = {{Vathachira}, Irin Babu and {Hillman}, Yael and {Kashi}, Amit},
        title = "{The Role of Binary Configuration in Shaping Nova Evolution via Wind Accretion in Symbiotic Systems}",
      journal = {\apj},
         year = 2026,
        month = feb,
       volume = {997},
       number = {2},
          eid = {278},
        pages = {278},
          doi = {10.3847/1538-4357/ae27c7},
archivePrefix = {arXiv},
       eprint = {2511.14596},
 primaryClass = {astro-ph.SR},
       adsurl = {https://ui.adsabs.harvard.edu/abs/2026ApJ...997..278V}
}

@ARTICLE{Wallstrom2025,
       author = {{Wallstr{\"o}m}, S.~H.~J. and {Scicluna}, P. and {Srinivasan}, S. and {Wouterloot}, J.~G.~A. and {McDonald}, I. and {Decock}, L. and {Wijshoff}, M. and {Chen}, R. and {Torres}, D. and {Umans}, L. and {Willebrords}, B. and {Kemper}, F. and {Rau}, G. and {Feng}, S. and {Jeste}, M. and {Kaminski}, T. and {Li}, D. and {Liu}, F.~C. and {Trejo-Cruz}, A. and {Chawner}, H. and {Goldman}, S. and {MacIsaac}, H. and {Tang}, J. and {Zeegers}, S.~T. and {Danilovich}, T. and {Matsuura}, M. and {Menten}, K.~M. and {van Loon}, J. Th and {Cami}, J. and {Clark}, C.~J.~R. and {Dharmawardena}, T.~E. and {Greaves}, J. and {He}, Jinhua and {Imai}, H. and {Jones}, O.~C. and {Kim}, H. and {Marshall}, J.~P. and {Shinnaga}, H. and {Wesson}, R.},
        title = "{The Nearby Evolved Stars Survey: III. First data release of JCMT CO-line observations}",
      journal = {\aap},
         year = 2025,
        month = dec,
       volume = {704},
          eid = {A276},
        pages = {A276},
          doi = {10.1051/0004-6361/202556298},
archivePrefix = {arXiv},
       eprint = {2510.14809},
 primaryClass = {astro-ph.GA},
       adsurl = {https://ui.adsabs.harvard.edu/abs/2025A&A...704A.276W}
}

@ARTICLE{Maes2021,
       author = {{Maes}, S. and {Homan}, W. and {Malfait}, J. and {Siess}, L. and {Bolte}, J. and {De Ceuster}, F. and {Decin}, L.},
        title = "{SPH modelling of companion-perturbed AGB outflows including a new morphology classification scheme}",
      journal = {\aap},
         year = 2021,
        month = sep,
       volume = {653},
          eid = {A25},
        pages = {A25},
          doi = {10.1051/0004-6361/202140823},
archivePrefix = {arXiv},
       eprint = {2107.00505},
 primaryClass = {astro-ph.SR},
       adsurl = {https://ui.adsabs.harvard.edu/abs/2021A&A...653A..25M}
}

%----------------------------------------------------------

\section{ACKNOWLEDGEMENTS}

The authors thank an anonymous referee for taking the time reviewing the original manuscript, and for comments and suggestions that improved the presentation of the results. JAT and DAVT thank support from UNAM DGAPA PAIIT project IN102324. RFM thanks UNAM DGAPA (Mexico) and SECIHTI (Mexico) for postdoc fellowships. This work made use of dedicated computing resources, which are developed and maintained by the personnel of the IRyA's Computing Department, Daniel Díaz-González, Miguel Espejel, Alfonso Ginori González, and Gilberto Zavala. All of them are gratefully acknowledged. This work has made an extensive use of NASA's Astrophysics Data System (ADS). 

\begin{appendices}
\setcounter{table}{0}
\renewcommand{\thetable}{A\arabic{table}}

\begin{table*}[t]
\begin{center}
%\footnotesize
\caption{Binary and orbital parameters estimated for symbiotic systems. The columns list the mass of the primary (donor) $m_1$, mass of the secondary (accretor) $m_2$, orbital separation $a$, orbital velocity $v_\mathrm{o}$, orbital period $T$, dimensionless mass ratio $q=m_2/(m_1+m_2)$, dimensionless velocity parameter $w = v_\mathrm{w}/v_\mathrm{o}$, mass accretion efficiencies $\eta_\mathrm{TT}$, Roche Lobe radius $r_\mathrm{LB}$ and the red giant radius $R_\mathrm{RG}$. We assume a typical wind velocity of $v_\mathrm{w}=12\pm8$~km~s$^{-1}$.} 
\begin{tabular}{lcccccccccccc}
\hline
Object & $m_1$      & $m_2$       & $P$ & $e$ & $a$  &$v_\mathrm{o}$ & $q$ & $w$ & $\eta_\mathrm{TT}$  & $R_\mathrm{RL}$ & $R_\mathrm{RG}$\\
       & [M$_\odot$]& [M$_\odot$] & [yr] &  & (AU)&  [km~s$^{-1}$]   &   &  &   & [AU] & [AU] \\
\hline
$o$ Cet &  2.00 &  0.65 & 497.90&  0.16 &87.3 & 5.2  & 0.25 & 2.30 & $1.6\times10^{-3}$ & 42.1 & 1.9 \\
AG Dra  &  1.20 &  0.50 &  1.51 &  0.0  &1.6  & 31.1  & 0.29 & 0.38 & $6.4\times10^{-2}$  & 0.7 & 0.16 \\
AG Peg  &  2.60 &  0.65 &  2.23 &  0.12 &2.5  & 33.8  & 0.20 & 0.35 & $3.1\times10^{-2}$ & 1.3 & 0.4 \\
BX Mon  &  3.70 &  0.55 &  3.78 &  0.44 &3.9  & 31.0  & 0.13 & 0.39 & $1.3\times10^{-2}$ & 2.2 & 0.7 \\
CH Cyg  &  2.20 &  0.56 & 15.58 &  0.12 &8.8  & 16.8  & 0.20 & 0.71 & $1.8\times10^{-2}$ & 4.4 & 1.3 \\
ER Del  &  3.00 &  0.70 &  5.72 &  0.17 &5.0  & 25.8  & 0.19 & 0.47 & $2.4\times10^{-2}$ & 2.5 & 0.5 \\
HD 330036 & 4.46 & 0.54 & 4.59  & \dots &4.8   & 30.7  & 0.11 & 0.39 & $9.1\times10^{-3}$ & 2.7 & 0.1 \\
IV Vir    & 0.90 & 0.42 & 0.77  & 0.0   &0.9   & 35.7  & 0.32 & 0.34 & $8.2\times10^{-2}$ & 0.4 & 0.1 \\
LT Del    & 1.00 & 0.57 & 1.24  & 0.40  &1.4   & 31.9  & 0.36 & 0.38 & $9.9\times10^{-2}$ & 0.6 & 0.14 \\
PU Vul  &  1.00 &  0.50 & 13.42 & 0.16  &6.5  & 14.4  & 0.33 & 0.83 & $3.8\times10^{-2}$ & 2.9 & 0.9 \\
R Aqr   &  1.00 &  0.70 & 42.40 & 0.45  &14.6 & 10.2  & 0.41 & 1.18 & $2.9\times10^{-2}$ & 6.0 & 1.4 \\
St 2-22   & 2.80 & 0.80 & 2.51  & 0.16  &2.9   & 33.6  & 0.22 & 0.36 & $3.8\times10^{-2}$ & 1.4 & 0.4 \\
TX CVn  &  3.50 &  0.40 &  0.55 & 0.56  &1.1  & 57.4  & 0.10 & 0.21 & $9.2\times10^{-3}$ & 0.6 & 0.1 \\
V471 Per  & 2.30 & 0.80 & 17.00 & \dots &9.4   & 16.4  & 0.26 & 0.73 & $2.3\times10^{-2}$ & 4.6 & 0.05 \\
V934 Her  & 1.60 & 1.35 & 12.02 & 0.35  &7.6   & 18.7  & 0.46 & 0.64 & $1.0\times10^{-1}$ & 3.0 & 0.4 \\
\hline
\end{tabular}
\label{tab:objects}
\end{center}
\end{table*}

\section{Observed Symbiotic Systems}
\label{sec:app}

Table~\ref{tab:objects} lists the properties of various symbiotic systems retrieved from the {\it New Online Database of Symbiotic Variables}\footnote{\url{https://sirrah.troja.mff.cuni.cz/~merc/nodsv/}} \citep{Merc2019}. This table includes the object name, the masses of the stellar components ($m_1$ and $m_2$), the binary period $P$, eccentricity (e), the estimated orbital separation ($a$), and the orbital velocity ($v_\mathrm{o}$). Furthermore, it provides the reduced mass ratio $q$, the dimensionless velocity parameter $w$, the estimated mass accretion efficiency $\eta_\mathrm{TT}$ obtained from Equation~(\ref{eq:eta_TT}), the Roche Lobe radius ($R_\mathrm{RL}$), and the radius of the late-type companion ($R_1 = R_\mathrm{RG}$).

For our analysis, we only selected symbiotic systems where the estimated Roche Lobe radius is at least three times larger than the radius of the late-type component ($R_\mathrm{RL}\geq 3\,R_\mathrm{RG}$). This criterion ensures that the systems are operating within the wind-accretion regime rather than being dominated by Wind Roche Lobe Overflow. We note that a similar selection condition was adopted by \citet{Maldonado2025}; consequently, the objects listed in Table~\ref{tab:objects} are the same as those analyzed in that work.

We note that the eccentricity values are listed when ever available at the {\it New Online Database of Symbiotic Variables}. We particularly note the case of R Aqr from which we list the value of $e=0.45$ from \citet{Alcolea2023}.

\end{appendices}

\end{document}